\documentstyle[aps,prd,psfig]{revtex}
\begin{document}
\renewcommand{\thesection}{\arabic{section}}
\renewcommand{\thesubsection}{\arabic{subsection}}

\title{Chiral Symmetry Breaking in Presence of Strong Quantizing
Magnetic Fields-\\ A Nambu-Jona-Lasinio Model with Semi-Classical Approximation}
\author{ Sutapa Ghosh$^{a)}$, Soma Mandal$^{a)}$ andSomenath Chakrabarty$^{a),b
)}${\thanks{E-Mail: somenath@klyuniv.ernet.in}}}
\address{$^{a)}$Department of Physics, University of Kalyani, Kalyani 741 235,
India and $^{b)}$Inter-University Centre for Astronomy and Astrophysics, Post 
Bag 4, Ganeshkhind, Pune 411 007, India}
%\date{\today}
\maketitle
\begin{abstract}
The breaking of chiral symmetry of light quarks at zero temperature in presence
of strong quantizing magnetic field is studied using Nambu-Jona-Lasinio (NJL) 
model with Thomas-Fermi type semi-classical formalism. It is found that the 
dynamically generated light quark mass can never become zero if the Landau 
levels are populated and increases with the increase of magnetic field strength.
\end{abstract}
The theoretical investigation of properties of compact stellar objects in 
presence of strong quantizing magnetic field have gotten a new life after the 
recent discovery of a few magnetars \cite{R1,R2,R3,R4}. These stellar objects 
are believed to be strongly magnetized young neutron stars. The surface magnetic
fields are observed to be $\geq 10^{15}$G. Then it is quite possible that the 
fields at the core region may go up to $10^{18}$G. The exact source of this 
strong magnetic field is of course yet to be known. These objects are
also supposed to be the possible sources of anomalous X-ray and soft gamma 
emissions (AXP and SGR). If the magnetic field is really so strong, in 
particular at the core region, they must affect most of the important physical 
properties of such stellar objects and also the rates / cross-sections of 
elementary processes, e.g., weak and electromagnetic decays / reactions taking 
place at the core region should also change significantly.

The strong magnetic field affects the equation of state of dense neutron
star matter. As a consequence the gross-properties of neutron stars 
\cite{R5,R6,R7,R8}, e.g., mass-radius relation, moment of inertia,
rotational frequency etc. should change significantly. In the case of
compact neutron stars, the phase transition from neutron matter to quark 
matter which may occur at the core region is also affected by strong quantizing
magnetic field.  It has been shown that a first order phase transition initiated
by the nucleation of quark matter droplets is absolutely forbidden if the 
magnetic field strength $\sim 10^{15}$G at the core region \cite{R9,R10}. 
However, a second order phase transition is allowed, provided the magnetic 
field strength $<10^{20}$G.   This is of course too high to achieve at the core
region.

The elementary processes, in particular, the weak and the electromagnetic
decays/reactions taking place at the core region of a neutron star are strongly 
affected by such ultra-strong magnetic fields \cite{R11,R12}. Since the cooling 
of neutron stars are mainly controlled by neutrino/anti-neutrino emission, the 
presence of strong quantizing magnetic field should affect the thermal history 
of strongly magnetized neutron stars.  Further, the electrical conductivity of 
neutron star matter which directly controls the evolution of neutron star 
magnetic field will also change significantly \cite{R12}.

Similar to the study of quark-hadron deconfinement transition inside neutron 
star core in presence of strong quantizing magnetic field, some investigations 
have also been done on the effect of ultra-strong magnetic field on chiral 
symmetry breaking. In those studies,  quantum field theoretic formalism were 
mainly used \cite{R13,R14,R15,R16,R17,R18,R19}. In particula, in the
reference \cite{Rina}, chiral symmetry violation is studied with NJL
model with quantum field theoretic approach in presence of strong
magnetic field and gravitational field and the calculations were done
exactly.

In this article we shall study the effect of strong quantizing magnetic
field on the chiral properties of light quark matter system with the help
of NJL model following semi-classical Thomas-Fermi type mean field approach.
Now in NJL model, there is no in-built mechanism of color confinement, however,
it can produce two chirally distinct phases- appropriate for confined quark 
matter within the bag (not necessarily tiny hadronic bag) and the matter outside
the bag. These phases are also known as the Wigner phase and spontaneously 
broken chiral phase respectively. Therefore, if one re-formulates the NJL model
in presence of strong quantizing magnetic field, it is quite possible to obtain
the effect of quantizing magnetic field on these two chirally distinct phases 
and hence obtain the effect of magnetic field on chiral symmetry breaking. 
Further, it is also possible to obtain bag pressure from the difference of 
vacuum energy densities of these two phases and hence its variation with strong
magnetic field. Assuming that the confinement and spontaneously broken chiral 
symmetry are synonymous, Bhaduri et. al. obtained some estimate of bag constant 
from the difference of energy densities \cite{R20}.  In the present article we 
shall modify these original calculations of Bhaduri et. al. \cite{R20} and 
Provid${\hat{\rm{e}}}$ncia et. al. \cite{R21} to study the breaking of chiral 
symmetry of light quarks in presence of strong magnetic fields and show that 
the chiral symmetry always remains broken in presence of strong quantizing 
magnetic field if the Landau levels for quarks are populated.  Our study is 
basically an application of the  formalism recently developed to study the 
equation of state of dense fermionic matter of astrophysical interest in 
presence of strong quantizing magnetic field \cite{R22}.

We start with the density matrix $\rho(x,x^\prime)$, given by
\begin{equation}
\rho(x,x^\prime)=\sum_{{\rm{spin}},p} \psi(x)\psi^\dagger(x^\prime) 
\theta(\Lambda-\mid p_z\mid)
\end{equation}
where $ \psi $ and $ \psi^\dagger $ are respectively the negative energy Dirac 
spinor and the corresponding adjoint, satisfy the equation
\begin{equation}
h\psi=E_-\psi
\end{equation}
(and similarly for $ \psi^\dagger $) with the single particle Hamiltonian
\begin{equation}
h=\gamma_5 \vec \Sigma. (\vec p-q_f \vec A) +\beta m
\end{equation}
with
\begin{equation}
\vec \Sigma=\left ( \begin{array}{lr} \vec \sigma  & 0 \\ 0  & \vec \sigma\\
\end{array}
\right ),
\end{equation}
$\gamma_5$ and $\beta$ are the usual Dirac matrices, $\Lambda$ is the
ultra-violet cut off in the momentum integral over $p_z$ and $\vec A$ is
the electromagnetic field three vector corresponding to the external
constant magnetic field of strength $B_m$ along $z$-axis.  Here the light 
quark mass $m$ is assumed to be generated dynamically. Now in presence of 
strong quantizing magnetic field along $z$-direction, the up and down spin 
negative energy spinors are therefore given by
\begin{equation}
\psi(x)=\frac{1}{(L_yL_z)^{1/2}}\exp[i(E_\nu t-p_yy-p_zz)]v_-^{(\uparrow,
\downarrow)}
\end{equation}
where
\begin{equation}
v_-^{(\uparrow)}=\frac{1}{[2E_-(E_--m)]^{1/2}}\left ( \begin{array}{c}
p_zI_\nu\\ -i(2\nu q_fB_m)^{1/2} I_{\nu-1}\\ (E_--m)I_\nu \\ 0 \end{array}
\right )
\end{equation}
and
\begin{equation}
v_-^{(\downarrow)}=\frac{1}{[2E_-(E_--m)]^{1/2}}\left ( \begin{array}{c}
i(2\nu q_fB_m)^{1/2} I_\nu\\ -p_zI_{\nu-1}\\ 0\\
(E_--m)I_{\nu-1} \end{array}
\right )
\end{equation}
where $E_-=-(p_z^2+m^2+2\nu q_fB_m)^{1/2}=-E_\nu$, is the single particle 
energy eigen value, $\nu=0,1,2...$, are the Landau quantum numbers, $q_f$ is 
the magnitude of the charge carried by $f$th flavor and
\begin{equation}
I_\nu=\left (\frac{q_fB_m}{\pi}\right )^{1/4}\frac{1}{(\nu !)^{1/2}}2^{-\nu/2}
\exp \left [{-\frac{1}{2}q_fB_m\left (x-\frac{p_y}{q_fB_m} \right )^2}\right 
] H_\nu \left [(q_fB_m)^{1/2}\left (x-\frac{p_y}{q_fB_m} \right) \right ]
\end{equation}
with $H_\nu$ is the well known Hermite polynomial of order $\nu$, and $L_y$, 
$L_z$ are respectively length scales along $Y$ and $Z$ directions. Now it can 
very easily be shown that $\nu=0$ state is singly degenerate, whereas all other 
states are doubly degenerate. We now express the density matrix, as the modified
version of Wigner transform in presence of strong quantizing magnetic field, 
in the following form:
\begin{equation}
\rho(x,x^\prime)=\sum \rho(x,x^\prime,p_y,p_z,\nu)
\exp[i\{(t-t^\prime)E_-- (y-y^\prime)p_y- (z-z^\prime)p_z\}]
\end{equation}
where the sum is over the momentum components $p_y$, $p_z$ and the Landau 
quantum number $\nu$. Since the momentum variables are continuous, the sum over
momentum components will be replaced by the corresponding integrals. Then  
using the negative energy up and down spin Dirac spinors, we have
\begin{equation}
\rho(x,x^\prime,p_y,p_z,\nu)=\frac{1}{2E_-}[E_-
A-p_z\gamma_z\gamma_0A +m\gamma_0 A-p_\perp \gamma_y\gamma_0 B]
\theta(\Lambda-\mid p_z\mid)
\end{equation}
where the matrices $A$ and $B$ are given by
\begin{equation}
A=\left ( \begin{array}{l c c r}I_\nu I_\nu^\prime &0&0&0 \\
0 & I_{\nu-1}I_{\nu-1}^\prime &0 &0 \\
0 & 0 &I_\nu I_\nu^\prime &0 \\
0 & 0 & 0 &I_{\nu-1}I_{\nu-1}^\prime \\
\end{array}
\right )
\end{equation}
\begin{equation}
B= \left ( \begin{array}{l c c r}I_{\nu-1} I_\nu^\prime &0&0&0 \\
0 & I_\nu I_{\nu-1}^\prime &0 &0 \\
0 & 0 &I_{\nu-1} I_\nu^\prime &0 \\
0 & 0 & 0 &I_\nu I_{\nu-1}^\prime \\
\end{array}
\right )
\end{equation}
where the primes indicate the functions of $x'$. Now in the evaluation of
vacuum energy, we have noticed that it would be more convenient to define a
quantity $\mu_f$, similar to the chemical potential for the $f$th flavor
in a multi-quark statistical system in presence of strong quantizing magnetic 
field. Then it is very easy to write
\begin{equation}
\Lambda=\left ({\mu_f}^2-m^2-2\nu q_f B_m \right)^{1/2}
\end{equation}
Since $\Lambda>0$, it is also possible to express the upper limit of $\nu$,
which is the maximum value of Landau quantum number of the levels occupied by 
$f$th flavor, in terms of $\mu_f$, $m$, $q_f$ and $B_m$ and is given by
\begin{equation}
\nu_{\rm{max}}^{(f)}=\left [ \frac{\mu_f^2-m^2}{2q_fB_m}\right ]
\end{equation}
where $[~]$ indicates the nearest integer but less than the actual number.
Now to obtain the energy density of the vacuum, we consider the NJL (chiral) 
Hamiltonian, given by
\begin{eqnarray}
H&=& \sum_{i=i}^N t(i)+\frac{1}{2} \sum_{i\neq j} V(i,j)\\
&=& \sum_{i=1}^N\gamma_5(i)\vec \Sigma(i).(\vec p_i- q_f\vec A)-\frac{1}{2}
\sum_{i\neq j} \delta(\vec x_i-\vec x_j)[\beta(i)\beta(j)- \beta(i)
\gamma_5(i) \beta(j) \gamma_5(j)]
\end{eqnarray}
Assuming the magnetic field $B_m$ along $z$-direction and is constant, we can 
choose the gauge $A^\mu\equiv (0,0,xB,0)$. The energy of the vacuum is then 
given by
\begin{equation}
\epsilon_v=\sum_{p_{1_z},\nu_1}\int dx_1 tr_1[\{\gamma_5\vec \Sigma.
(\vec p_1-q_f \vec A)\}\rho_{p_1}]+ \epsilon_v^{(I)} 
\end{equation}
where $\rho_{p_1}$ is given by eqn.(10) and $\epsilon_v^{(I)}$ indicates
the interaction term, including the exchange interaction. To evaluate the 
vacuum energy, we first calculate the first term of eqn.(17). This quantity is
proportional to the trace defined as $Tr(\rho h)$, can easily be evaluated by 
using $\rho$ from eqn.(10) and the single particle Hamiltonian $h$ from 
eqn.(16). Now using the orthonormality relations for the Hermite polynomials at
the time of evaluation of integral over $dx$ and also using the anti-commutation
relations of $\gamma$-matrices, we have the first term at zero temperature
\begin{equation}
\epsilon_v^{(0)}=2N_c\sum_{f=u,d}\frac{q_fB_m}{2\pi^2}
\sum_{\nu=0}^{\nu_{\rm{max}}} (2-\delta_{\nu 0})
\int_0^\Lambda  dp_z \frac{\vec p^2}{E_-}
\end{equation}
where $\vec p^2=p_z^2+2\nu q_fB_m$, $N_c=3$, the number of colors, and $E_-
=-E_\nu$.

In the evaluation of all the traces in this paper we have used the
following important relation:
\begin{equation}
{\rm{Tr}}(\gamma^\mu \gamma^\nu A_1A_2..B_1B_2..)=Tr(A_1A_2..B_1B_2..)
g^{\mu\nu},
\end{equation}
\begin{equation}
{\rm{Tr}}(\gamma^\mu\gamma^\nu\gamma^\lambda\gamma^\sigma
A_1A_2..B_1B_2..)=Tr(A_1A_2..B_1B_2..)(g^{\mu \nu}
g^{\sigma \lambda}-g^{\mu \lambda}g^{\nu \sigma}+g^{\mu \lambda}g^{\nu
\sigma}), 
\end{equation}
${\rm{Tr}}$(product of odd $\gamma$s with  $A$ and/or $B)=0$  etc. The other
interesting aspects of $A$ and $B$ matrices are:\\
i) $k_{1\mu}k^{2\mu}{\rm{Tr}}(A_1A_2)= (E_1E_2-k_{1z}k_{2z}){\rm{Tr}}(A_1A_2)$\\
ii) $k_{1\mu}k^{2\mu}{\rm{Tr}}(B_1B_2)= \vec k_{1\perp}.\vec
k_{2\perp}{\rm{Tr}}(B_1B_2)$\\
iii) $k_{1\mu}k^{2\mu}{\rm{Tr}}(A_1B_2)= k_{1\mu}k^{2\mu}{\rm{Tr}}(B_1A_2)=0$\\
iv) $p_{1\mu}k^{1\mu}p_{2\nu}k^{2\nu}{\rm{Tr}}(A_1B_2)\neq 0=
(E_{\nu_1}E_{\nu_2^\prime}-p_{1z}k_{1z})\vec p_{2\perp}.\vec k_{2\perp}
{\rm{Tr}}(A_1B_2)$\\
These set of relations are very recently obtained by us \cite{R22}.
Since $\gamma$ matrices are traceless and both $A$ and $B$ matrices are 
diagonal with identical blocks, it is very easy to evaluate the above traces of 
the product of $\gamma$-matrices multiplied with any number of $A$ and/or $B$, 
from any side with any order. 

To evaluate the interaction term, we first consider the direct part which is
proportional to $Tr(\beta \rho_{p_1})Tr(\beta\rho_{p_2})$ and it is very
easy to show that $Tr(\beta \gamma_5\rho)=0$. Then using the orthonormality 
relations for Hermite polynomials and the anti-commutation relations for the 
$\gamma$-matrices, we have the direct term
\begin{equation}
V_{\rm{dir}}=-4gm^2[{\cal{V}}(\Lambda, m)]^2
\end{equation}
where
\begin{equation}
{\cal{V}}(\Lambda,m)=\frac{N_c}{2\pi^2} \sum_{f=u,d} e_fB_m 
\sum_{\nu=0}^{\nu_{\rm{max}}} (2-\delta_{\nu 0}) \int_0^\Lambda
\frac{dp_z}{(p_z^2+m_\nu^2)^2}
\end{equation}
where $m_\nu=(m^2+2\nu q_f B_m)^{1/2}$.

To evaluate the exchange term, we first calculate $Tr((\beta\rho_{p_1})
(\beta\rho_{p_2}))$. Now
\begin{equation}
\beta\rho_p=\frac{1}{2E_-}[E_-\beta A+p_zA\gamma_z +mA -p_\perp
B\gamma_y]
\end{equation}
Then at the time of integration over $dx_1$ and $dx_2$ if one uses the
orthonormality relations for Hermite polynomials, the above trace is given by
\begin{equation}
\left [ 1+\frac{p_{1z}p_{2z}}{E_1E_2}+\frac{m^2}{E_1E_2}\right ]
\end{equation}
where both $E_1$ and $E_2$ are negative. Then in the energy contribution, after
integrating over $p_{1z}$ and $p_{2z}$, the first term gives 
\begin{equation}
\left ( \frac{N_c}{2\pi^2}\sum_fq_fB_m\sum_\nu (2-\delta_{\nu 0}) \Lambda \right
)^2
\end{equation}
Similarly the contribution from second term is given by
\begin{equation}
\left ( \frac{N_c}{2\pi^2}\sum_fq_fB_m\sum_\nu (2-\delta_{\nu 0})
(\Lambda^2+m_\nu^2)^{1/2}\right)^2
\end{equation}
and finally, the third term is gives
\begin{equation}
m^2\left ( \frac{N_c}{2\pi^2}\sum_fq_fB_m\sum_\nu \gamma_\nu
\ln\left [ \frac{\Lambda +(\Lambda^2+m_\nu^2)^{1/2}}{m_\nu}\right ]\right)^2
\end{equation}
To obtain the next term in the exchange part, we evaluate the trace
$Tr((\beta\gamma_5\rho_{p_1})(\beta\gamma_5\rho_{p_2}))$, which unlike
the direct case, gives non-zero contribution. Using the anticommutation 
relations of $\gamma$-matrices and as usual with the help of ortonormality 
relations for Hermite polynomials, we have the above trace
\begin{equation}
-\left [ 1+\frac{p_{1z}p_{2z}}{E_1E_2}+ \frac{m^2}{E_1E_2} +m\left (
\frac{1}{E_1} +\frac{1}{E_2} \right ) \right ]
\end{equation}
The contribution to the interaction energy will again be obtained if we
integrate over $p_{1z}$ and $p_{2z}$. Then the first term is given by
\begin{equation}
\left ( \frac{N_c}{2\pi^2}\sum_fq_fB_m\sum_\nu (2-\delta_{\nu 0})
\Lambda\right)^2
\end{equation}
The second term is given by
\begin{equation}
\left ( \frac{N_c}{2\pi^2}\sum_fq_fB_m\sum_\nu (2-\delta_{\nu 0})
(\Lambda^2+m_\nu^2)^{1/2}\right)^2
\end{equation}
The third term is given by
\begin{equation}
m^2\left ( \frac{N_c}{2\pi^2}\sum_fq_fB_m\sum_\nu (2-\delta_{\nu 0})
\ln\left [ \frac{\Lambda +(\Lambda^2+m_\nu^2)^{1/2}}{m_\nu}\right ]\right)^2
\end{equation}
and finally the fourth and fifth terms, which  are identical, given by
\begin{equation}
m\left ( \frac{N_c}{2\pi^2}\sum_fq_fB_m\sum_\nu (2-\delta_{\nu 0})
\ln\left [ \frac{\Lambda +(\Lambda^2+m_\nu^2)^{1/2}}{m_\nu}\right ]\right)
\left ( \frac{N_c}{2\pi^2}\sum_fq_fB_m\sum_\nu \gamma_\nu
\Lambda\right)
\end{equation}

Then combining all these terms we finally obtain the vacuum energy
density. Since the mass $m$, which is assumed to be same for both $u$
and $d$ quarks, is generated dynamically, we obtain this quantity by
minimizing the total vacuum energy density with respect to $m$, i.e.,
by putting $d\epsilon_v/dm=0$. Simplifying this non-linear equation, we
finally get
\begin{equation}
\frac{d\epsilon_v}{dm}=-P+2gQR=0
\end{equation}
where
\begin{eqnarray}
P&=&\frac{N_c}{2\pi^2}\sum_fq_fB_m\sum_\nu(2-\delta_{\nu 0})\left [
 \frac{2 m^3 \Lambda}{m_\nu^2 }\frac{1}{(\Lambda^2+m_\nu^2)^{1/2}}
-2mX\right]\\
Q&=&\frac{N_c}{2\pi^2}\sum_fq_fB_m\sum_\nu(2-\delta_{\nu 0})\left [X-
\frac{m^2}{m_\nu^2} \frac{\Lambda}{(\Lambda^2+m_\nu^2)^{1/2}} \right ]\\
R&=&\frac{N_c}{2\pi^2}\sum_fq_fB_m\sum_\nu(2-\delta_{\nu 0})\left [\Lambda- 4mX
\right ]
\end{eqnarray}
with
\begin{equation}
X=\ln\left [ \frac{\Lambda +(\Lambda^2+m_\nu^2)^{1/2}}{m_\nu}\right ]
\end{equation}

It is therefore obvious from eqn.(33) that the trivial solution $m=0$ is
not possible in this particular situation. On the other hand in a
non-magnetic case, eqn.(33) reduces to the well known gap equation, given by
\begin{equation}
m=4g{\cal{V}} m 
\end{equation}
where ${\cal{V}}$ is the overall contribution of interaction terms. Hence
it is obvious that $m=0$, the trivial solution exists in this
non-magnetic or the conventional scenario, investigated by Bhaduri et.
al. \cite{R20}. The phase with $m=0$ is the Wigner phase and $m\neq 0$ is the 
so called Goldstone phase, which further gives
\begin{equation}
4g{\cal{V}}=1
\end{equation}
which is nothing but the well known gap equation used in BCS theory.

The non-existence of trivial solution indicates the spontaneously broken
chiral symmetry in presence of strong quantizing magnetic field. Therefore 
as soon as the Landau levels are populated for light quarks in
presence of external magnetic field, the chiral symmetry gets broken,
the quarks become massive and the mass $m$ (assumed to be same for both
$u$ and $d$ quarks) is generated dynamically. As mentioned earlier, that such
investigations were done with quantum field theory. On the other hand the
approach followed by us in this article is in some sense semi-classical and 
have not been done before.

Therefore we may conclude that the Wigner phase does not exist in the case of 
relativistic Landau dia-magnetic system. Further, if the deconfinement 
transition and restoration of chiral symmetry occur simultaneously, then it 
puts a big question mark whether the idea of bag model is applicable at all in
presence of strong quantizing magnetic field. 

To illustrate the variation of dynamical quark mass with magnetic field, we 
consider the relation
\begin{equation}
m_\pi^2=-\frac{m_0}{f_\pi^2}<\psi \bar \psi>
\end{equation}
where $m_\pi$ is the pion mass, $m_0$ is the quark current mass and $f_\pi$ is 
the pion decay constant. Using the spinor solutions given by eqns.(6) and (7) 
we get
\begin{equation}
m_\pi^2=\frac{2m_0m}{f_\pi^2}~\frac{N_c}{2\pi^2}\sum_{f=u,d} \sum_{\nu
=0}^{\nu_{\rm{max}}} (2-\delta_{\nu 0}) \ln\left
[\frac{\Lambda+(\Lambda^2+m_\nu^2)^{1/2}}{m_\nu}\right ]
\end{equation}
We have now solved the eqns.(33) and (41) numerically to obtain $\Lambda$ and 
$m$ for various values of magnetic field. We have considered the following sets
of numerical values for the parameters. The current quark mass $m_0=10$MeV, 
pion mass $m_\pi=140$MeV, pion decay constant $f_\pi=93$MeV, coupling constant 
$g=10$GeV$^{-2}$ and electron mass $m_e=0.5$MeV. In fig.(1) we shown the 
variation of dynamically generated quark mass with the strength of magnetic 
field. As it is evident that the dynamical quark mass never goes to zero and 
diverges beyond $B_m\approx 10^{17}$G.
%---------------   FIGURES ---------------------------------
\begin{figure}
\psfig{figure=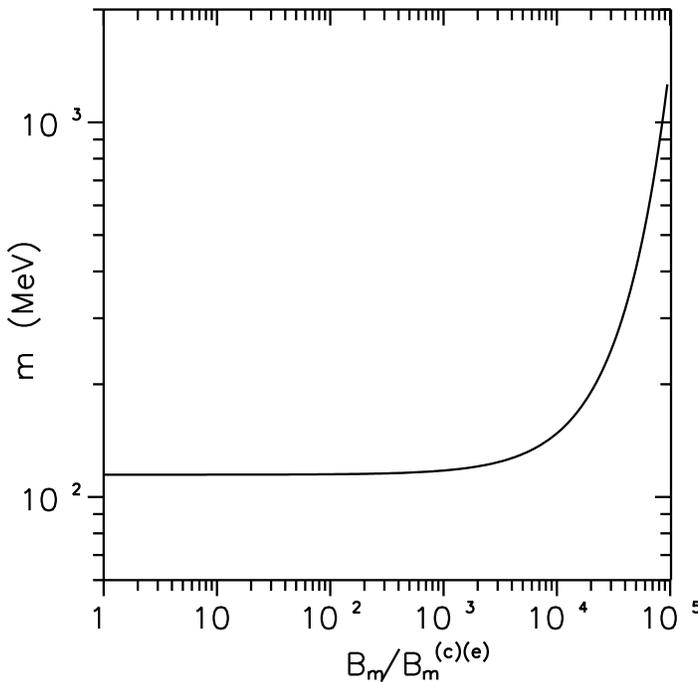,height=0.5\linewidth}
\caption{The variation of dynamically generated quark mass with the strength of
magnetic field (expressed in terms of $B_m^{(c)(e)}=4.4\times 10^{13}$G.)}
\end{figure}

\end{document}